\documentclass[prb, twocolumn, longbibliography, nofootinbib]{revtex4-2}

\usepackage{graphicx} 
\usepackage{amsmath}
\usepackage{amsfonts}
\usepackage{amssymb}
\usepackage[dvipsnames]{xcolor}
\usepackage[colorlinks=true]{hyperref}
\usepackage{graphicx}
\graphicspath{{doodles}}
\usepackage{tikz-cd}
\usepackage{subfigure}
\usepackage{bbold}

\hypersetup{linkcolor = BrickRed, citecolor= ForestGreen, urlcolor= RoyalBlue}
\definecolor{myred}{HTML}{ed3624}
\definecolor{mygreen}{HTML}{13b205}
\definecolor{myblue}{HTML}{0379ee}
\definecolor{mypink}{HTML}{ff6ed8}
\definecolor{myyellow}{HTML}{fbe601}

\makeatletter
\newcommand*\rel@kern[1]{\kern#1\dimexpr\macc@kerna}
\newcommand*\widebar[1]{%
  \begingroup
  \def\mathaccent##1##2{%
    \rel@kern{0.8}%
    \overline{\rel@kern{-0.8}\macc@nucleus\rel@kern{0.2}}%
    \rel@kern{-0.2}%
  }%
  \macc@depth\@ne
  \let\math@bgroup\@empty \let\math@egroup\macc@set@skewchar
  \mathsurround\z@ \frozen@everymath{\mathgroup\macc@group\relax}%
  \macc@set@skewchar\relax
  \let\mathaccentV\macc@nested@a
  \macc@nested@a\relax111{#1}%
  \endgroup
}
\makeatother

\renewcommand{\cal}[1]{\mathcal{#1}}

\renewcommand{\S}{\cal{S}}
\newcommand{\G}{\cal{G}}
\newcommand{\C}{\cal{C}}
\newcommand{\Z}{\cal{Z}}
\newcommand{\pardag}{\partial^\dagger\!}
\newcommand{\inlinegraphic}[1]{\vcenter{\hbox{\includegraphics{#1}}}}
\newcommand{\Para}{\text{Para}}
\newcommand{\RBH}{\text{RBH}}
\newcommand{\TCE}{\text{TCE}}
\newcommand{\TCF}{\text{TCF}}
\newcommand{\TTC}{\text{2TC}}
\newcommand{\ITC}{\text{ITC}}
\newcommand{\U}{U_{CX}}
\renewcommand{\hat}{\widehat}
\newcommand{\barelogical}[1]{\widebar{#1}}
\newcommand{\dressedlogical}[1]{\widetilde{#1}}

\begin{document}

\title{Single-Shot Quantum Error Correction in Intertwined Toric Codes}
\author{Charles Stahl}
\affiliation{Department of Physics, University of Colorado, Boulder, CO 80309, USA}
\date{August 27, 2024}

\begin{abstract}

We construct a subsystem code in three dimensions that exhibits single-shot error correction in a user-friendly and transparent way. As this code is a subsystem version of coupled toric codes, we call it the intertwined toric code (ITC). Although previous codes share the property of single-shot error correction, the ITC is distinguished by its physically motivated origin, geometrically straightforward logical operators and errors, and a simple phase diagram. The code arises from 3d stabilizer toric codes in a way that emphasizes the physical origin of the single-shot property. In particular, starting with two copies of the 3d toric code, we add check operators that provide for the confinement of pointlike excitations without condensing the loop excitations. Geometrically, the bare and dressed logical operators in the ITC derive from logical operators in the underlying toric codes, creating a clear relationship between errors and measurement outcomes. The syndromes of the ITC resemble the syndromes of the single-shot code by Kubica and Vasmer, allowing us to use their decoding schemes. We also extract the phase diagram corresponding to ITC and show that it contains the phases found in the Kubica-Vasmer code. Finally, we suggest various connections to Walker-Wang models and measurement-based quantum computation.

\end{abstract}

\maketitle

\section{Introduction} \label{sec:intro}

Single-shot quantum error-correcting codes provide for a streamlined error-correction protocol, in which a single round of imperfect measurements suffices to correct errors~\cite{BombinSingleShot}. The two-dimensional (2d) toric code~\cite{KitaevFaultTolerant} does not realize this property because the pointlike excitations cannot be reliably located in a single round of faulty measurements~\cite{DennisTopological}. On the other hand, the excitations in the 4d toric code are all extended excitations, allowing for single-shot error correction~\cite{DennisTopological}. The toric codes are all stabilizer codes, meaning that all operators in the code commute. In stabilizer codes, single-shot error correction is closely related to self-correction, suggesting that any single-shot stabilizer code is, in fact, self-correcting. No single-shot stabilizer codes have been found below 4d.

On the other hand, single-shot \emph{subsystem} codes do exist in 3d.
Subsystem codes generalize stabilizer codes by having a check group, which does not commute, and a stabilizer group, which does commute.
The gauge color code (GCC)~\cite{BombinGauge} was the first 3d subsystem code shown to be single-shot~\cite{BombinSingleShot}. Despite the code's remarkable properties, its complicated construction meant that no other examples were found until seven years later, by Kubica and Vasmer~\cite{KubicaVasmer}. Even in the Kubica-Vasmer code (KVC), the essential physical properties that provide for the success of single-shot error correction remain elusive. 

\begin{figure}[t]
    \centering
    \subfigure[\label{subfig:syndrome-toric}]{\includegraphics[width=.49\linewidth]{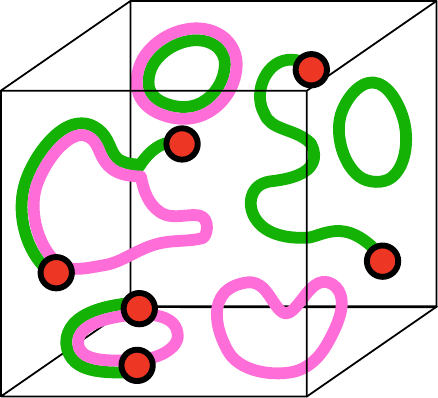}}
    \hfill
    \subfigure[\label{subfig:syndrome-color}]{\includegraphics[width=.49\linewidth]{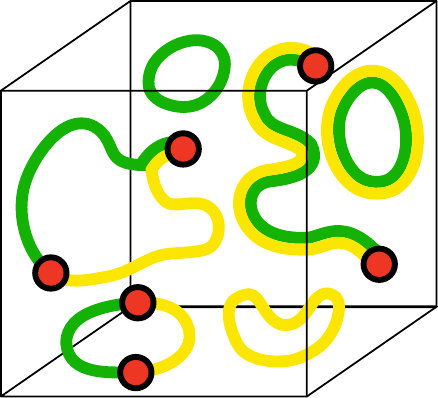}}
    \caption[Short description]{Characteristic measurement outcomes in the ITC. (a) In the underlying toric code, \textcolor{mygreen}{green} lines are truncated stringlike logical operators and \textcolor{mypink}{pink} lines are extended excitations. In the ITC, these objects are put on equal footing as lines of violated check operators~\eqref{eqn:checks-toric}. In both the toric code and the ITC, the \textcolor{myred}{red} dots are violated stabilizers. (b) With the alternative basis of check operators~\eqref{eqn:checks-color}, measurement outcomes consist of two colors of extended objects (\textcolor{mygreen}{green} and \textcolor{myyellow}{yellow} lines). These cannot end except on violated stabilizers, which must be the simultaneous endpoints of one line of each color.}
    \label{fig:syndrome}
\end{figure}

Here, we construct a single-shot 3d quantum error-correcting code that we call the intertwined toric code (ITC). This new code displays three main advantages: a transparent physical motivation and interpretation, geometrically simple logical operators, and a straightforward derivation of its phase diagram.
Physically, the ITC draws inspiration from the 3d toric code. Its check group consists of the toric code operators supplemented by single-site check operators. As a result, the ITC stabilizers detect the pointlike excitations and the check operators detect the stringlike operators that make those excitations, as depicted in Fig.~\ref{subfig:syndrome-toric}. This clarifies that, physically, the ITC confines the pointlike excitations without condensing the extended excitations---which is not possible in stabilizer codes. 

Measuring a different set of generators in the check group leads to check syndromes reminiscent of the KVC. In particular, the effective pictures of the syndromes in this second basis of the ITC contain two flavors of violated stabilizers connected by four flavors of check operators (whereas the GCC has eight stabilizers and 24 check operators). We illustrate one sector of this presentation of the ITC in Fig.~\ref{subfig:syndrome-color}. In this setting, the advantage of the ITC is the geometric simplicity of its logical operators: Membrane operators are defined on membranes and dual membranes on the lattice, and stringlike operators are defined on paths and dual paths on the lattice.
Furthermore, it is easy to find a partial logical operator that creates any given measurement outcome, as described at the end of Sec.~\ref{sub:local}.

Readers with a background in condensed matter may view subsystem codes as corresponding to families of Hamiltonians, with multiple gapped phases of matter separated by phase transitions. Such analyses exist for the GCC~\cite{KubicaYoshida} and the KVC~\cite{LiPhaseDiagram}. The ITC phase diagram is particularly easy to extract, and we find all of the phases of the KVC along with one additional phase.

As an aside, the subsystem code literature uses the term ``gauge operator" and ``gauge group" instead of ``check operator" and ``check group." To avoid confusion with the condensed-matter literature, where error-correcting codes often have interpretation in terms of gauge theories, we avoid using the word ``gauge operator" herein. Instead, we refer to the relevant operators as ``check operators" in the subsystem code, because one ``checks" the value of these operators during the error-correction procedure~\cite{HastingsHaah}.

The remainder of this paper is organized as follows. In Sec.~\ref{sec:build} we intertwine two toric codes, find a logical qubit, and describe the logical operators. We use a different set of check operators within the ITC in Sec.~\ref{sec:errors} to explore the possible measurement outcomes and outline the single-shot error correction procedure. In Sec.~\ref{sec:phase}, we describe the phase diagram and explore connections of the various phases to previous literature. We conclude in Sec.~\ref{sec:questions} with some exploratory connections to other constructions and with a discussion of some open questions.

\section{Intertwining toric codes} \label{sec:build}

To provide readers with some background, we start with a brief review of subsystem codes in Sec.~\ref{sub:subsystem}. Then, to define the ITC, we define the check group and stabilizer group on a cubic lattice with periodic boundary conditions in Sec.~\ref{sub:bulk}. Like the GCC and the KVC, the ITC does not encode any logical qubits on closed manifolds. To encode logical qubits we define boundary conditions in Sec.~\ref{sub:boundary}. The boundary conditions furnish the ITC with logical qubits, and we describe the logical operators on those qubits in Sec.~\ref{sub:logicals}.

\subsection{Brief review of subsystem codes} \label{sub:subsystem}

Recall that a stabilizer code consists of a stabilizer group $\S$, which is a commuting subgroup of the Pauli group that does not contain the global phase operator $-\mathbb{1}$. Elements of $\S$ are called stabilizers. A state is a code state if it satisfies all of the stabilizers, which means that it is a $+1$ eigenstate of those operators. Logical operators are elements of the centralizer of $\S$, $\Z(\S)$, which is the group of all Pauli operators that commute with $\S$.
These operators map code states to code states. Of course, every stabilizer is in $\Z(\S)$, but these are trivial logical operators. The nontrivial logical operators are elements of $\Z(\S) / \S$. Qubit errors are operators that do not commute with $\S$, and therefore take a code state out of the code space. A state not in the code space violates some stabilizers, meaning it is a $-1$ eigenstate of those stabilizers. 
A syndrome is the set of violated stabilizers in a given state, and an error causes a syndrome when it anticommutes with that set of stabilizers. 

Subsystem codes~\cite{PoulinSubsystem} provide more flexibility than stabilizer codes by including a check group $\G$ in addition to a stabilizer group. The check group does not need to commute and may contain global phase operators. Heuristically, we think of the stabilizer group as the center of $\G$, which is $\Z(\G) \cap \G$, or the group made up of operators in $\G$ that commute with all operators in $\G$. However, this may contain pure phases, so the stabilizer group of a subsystem code is actually
\begin{align}
\S = (\Z(\G) \cap \G) / i,
\end{align}
the center of $\G$ with pure phases removed. Since $\S \subset \G$, each stabilizer is a product of check operators, allowing one to measure check operators in order to infer the value of stabilizers.

Code states cannot satisfy all of the check operators because $\G$ does not commute. Instead, like in a stabilizer code, code states satisfy all of the stabilizers. Within the code space there are logical degrees of freedom (``logical qubits") and other degrees of freedom (``check qubits").
The check operators act on the check qubits. Logical operators still must commute with all stabilizers, but come in two flavors depending on the degrees of freedom on which they act. Bare logical operators act only on the logical qubits while dressed logical operators act on the check qubits as well. Bare logical operators must commute with all check operators, making them elements of $\Z(\G)$. Dressed logical operators need only commute with the stabilizers, making them elements of $\Z(\S)$. Automatically, $\G \subset \Z(\S)$ and $\S \subset \Z(\G)$, so that the nontrivial bare and dressed logical operators are elements of $\Z(\G)/\S$ and $\Z(\S)/\G$, respectively.

The error-correction procedure consists of measuring the check operators, so the measurement outcome is the set of violated check operators. From a measurement outcome one can infer the stabilizer syndrome (the set of violated stabilizers). Code states have trivial stabilizer syndromes but still must have some nontrivial measurement outcomes. An operator that anticommutes with some check operators changes the measurement outcome, while an operator that anticommutes with some stabilizers causes a syndrome. 
In this language, dressed logical operators create trivial stabilizer syndromes but change the measurement outcome, while bare logical operators create trivial stabilizer syndromes and do not change the measurement outcome.

Subsystem codes provide a number of advantages over stabilizer codes. As in this paper, subsystem codes may be single-shot in 3d, while no known examples of such stabilizer codes exist. Subsystem codes also allow for ``gauge fixing"~\cite{PoulinSubsystem} a procedure we explore in Sec.~\ref{sec:phase}. They may allow for measuring smaller operators; for example, the subsystem toric code in Ref.~\cite{BravyiSTC} has check operators that only act on three qubits. Some further information on subsystem codes may be found in Ref.~\cite{EllisonSubsystem}, which studies anyon theories in 2d subsystem codes, and Ref.~\cite{BombinGauge}.

\subsection{In the bulk} \label{sub:bulk}

To define the ITC, start with two copies of the 3d \emph{stabilizer} toric code, defined on a periodic cubic lattice. With qubits on edges $e$ and faces $f$, define $A$ operators
\begin{align}
A_v &= \prod_{e \in \pardag v} X_e = \inlinegraphic{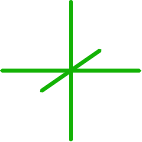},\\
A_c &= \prod_{f \in \partial c} Z_f = \inlinegraphic{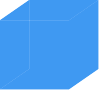},
\end{align}
for every vertex $v$ and cube $c$.
\textcolor{mygreen}{Green} and \textcolor{myblue}{blue} denote $X$ and $Z$ operators, respectively, and solid lines and shaded squares represent operators on edge qubits and face qubits, respectively. 
In addition, define $B$ operators
\begin{align}
B_f &= \prod_{e \in \partial f} Z_e = \inlinegraphic{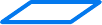},\\
B_e &= \prod_{f \in \pardag e} X_f      = \inlinegraphic{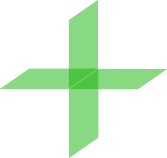},
\end{align}
and rotations thereof.
The operator $\partial$ is the boundary operator, so $\partial f$ is the four edges around a face $f$ and $\partial c$ is the six faces around a cube $c$. The operator $\pardag$ is the dual boundary, so $\pardag e$ is the four faces around an edge $e$ and $\pardag v$ is the six edges around a vertex $v$. These operators are dual, meaning that, for example, $f \in \pardag e \leftrightarrow e \in \partial f$

The stabilizers $A_v$ and $B_f$ together define the ordinary presentation of the 3d toric code, while $A_c$ and $B_e$ are related to them by interchanging edges with faces, cubes with vertices, and $X$ operators with $Z$ operators.
The stabilizer group
\begin{align}
\S_\text{2TC} = \left \langle A_v, B_f, A_c, B_e \right \rangle, \label{eqn:S-2TC}
\end{align}
which is the group generated by the terms inside the angled brackets, therefore defines two noninteracting 3d toric codes.
Violations of the $A$ stabilizers are isolated excitations, while violations of the $B$ stabilizers must form closed strings.
Thus, the excitations of this stabilizer code include two species of extended excitations and two species of pointlike excitations.

We now build the subsystem ITC from the stabilizer code $\S_\TTC$ by including the single-site operators $X_e$ and $Z_f$ on all edges and faces as check operators. The $B_e$ and $B_f$ operators anticommute with $Z_f$ and $X_e$, respectively, and are themselves demoted to check operators. The resulting ITC check group
\begin{align}
\cal{G}_\ITC &= \left \langle X_e, Z_f, B_e, B_f \right \rangle, \label{eqn:checks-toric}
\end{align}
includes anticommuting operators, and therefore also contains $-\mathbb{1}$. The $A$ operators still commute with all the check operators, so they define the stabilizer group
\begin{align}
\cal{S}_\ITC &= \Z(\cal{G}_\ITC) \cap \G_\ITC = \left \langle A_v, A_c \right\rangle. \label{eqn:stabilizers}
\end{align}
This construction shows that the ITC results from starting with the stabilizer toric code and ``confining" the pointlike excitations by including the single-site check operators $X_e$ and $Z_f$. 

This confinement procedure is worth discussing. In the realm of stabilizer codes, the only way to confine the pointlike excitations is to condense the extended excitations. Instead, subsystem codes allow for confinement without condensation. In the language of Ref.~\cite{EllisonSubsystem} we are ``gauging out" the extended excitations of the underlying toric codes, leaving the pointlike excitations as the only gapped excitations. This procedure is the essential process that allows for a physical interpretation of single-shot error correction in the ITC. The extended excitations do proliferate, but they remain detectable.

Note that $\G_\ITC$~\eqref{eqn:checks-toric} is invariant under the $\mathbb{Z}_2$ symmetry
\begin{align}
\U = \prod_{\langle e f\rangle} CX_{ef}, \label{eqn:UCX}
\end{align}
where $\langle e f \rangle$ are nearest neighbors and $CX_{ij}\, X_i\, CX_{ij} = X_iX_j$. The symmetry $\U$ maps $X_e \leftrightarrow X_eB_e$ and $Z_f \leftrightarrow Z_fB_f$, while leaving the $A$ and $B$ operators unchanged.

With periodic boundary conditions, there are also nonlocal stabilizers~\cite{EllisonSubsystem} in the ITC. First, note that arbitrary products of stabilizers look like membrane operators from the underlying stabilizer toric codes. Contractible membranes may be constructed this way. However, noncontractible membranes still commute with all the check operators and can be written as products of check operators. These operators are stabilizers by definition, but cannot be generated by the local stabilizers in~\eqref{eqn:stabilizers}. Thus, for every noncontractible membrane we may have to define a nonlocal $X_e$-type stabilizer and nonlocal $Z_f$-type stabilizer. 

On the three-torus (or any other manifold without boundary, regardless of genus) there are no logical qubits. We can show this by counting degrees of freedom. In an $L \times L \times L$ lattice, there are $3L^3$ edges and $3L^3$ faces. This means there are 
$N = 6L^3$ physical qubits and $3L^3$ of the $X_e$ (and $Z_f$) operators. There are $L^3$ distinct $A_v$ (and $A_c$) stabilizers, but only $L^3-1$ are independent due to the relation $\prod_v A_v = \mathbb{1}$ ($\prod_c A_c = \mathbb{1}$). There are also 6 nonlocal stabilizers. Similarly, there are $3L^3$ distinct $B_e$ (and $B_f$) operators, but $L^3+2$ relations between them~\cite{CastelnovoChamon}, for $2L^3-2$ independent $B_e$ (and $B_f$) operators. Altogether, there are
\begin{align}
K = N - \frac{1}{2} \left( \log_2|\G| + \log_2|\S| \right) = 0
\end{align}
logical qubits on the three-torus. In order to create logical qubits, we must define suitable boundary conditions.

\subsection{Constructing boundaries} \label{sub:boundary}

To define a logical qubit, let our cubic lattice have the topology $T^2\times I$, a cube with periodic boundary conditions in two directions. Suppose that the lattice is periodic in the front/back direction and the top/bottom direction. The boundary consists of two disjoint tori; on the right torus we impose the intertwined boundary conditions and on the left torus the trivial boundary conditions.  
In the \hyperref[sec:appendix]{Appendix} we construct the $e$- and $m$-condensed boundaries, giving rise to the ITC on a cube with no periodic boundary conditions.

The trivial (left) boundary follows simply from truncating our lattice in such a way that the boundary edges host qubits and boundary faces do not. We keep all the single-body $X_e$ and $Z_f$ check operators, along with any $B_e$ and $B_f$ check operators that remain four-body.
For a vertex $v$ and a boundary cube $c$ on the trivial boundary, this results in the stabilizers
\begin{align}
A_v &= \inlinegraphic{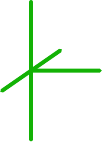}, \qquad A_c = \inlinegraphic{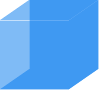}\label{eqn:triv-stabilizers}
\end{align}
each of which acts on five qubits. Note that we can use these to construct membrane operators that span the periodic directions, so that there are no independent nonlocal stabilizers given these boundary conditions.

To construct the intertwined (right) boundary, truncate the lattice in the same way, with qubits on boundary edges but not faces. On every boundary edge we replace the single-site $X_e$ check with the operators
\begin{align}
K_e = X_e X_{f(e)} =\inlinegraphic{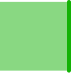}\;,\quad \inlinegraphic{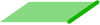} \label{eqn:twisted-checks}
\end{align}
where $f(e)$ is the unique nonboundary face containing $e$ in its boundary. The truncated $A_v$ operators are still stabilizers, but the truncated $A_c$ operators anticommute with some $K_e$ operators. We can construct valid stabilizers by dressing each naive $A_c$ with a $B_f$ operator to create
\begin{align}
A_v = \inlinegraphic{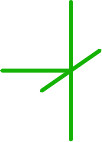},  \qquad A_c = \inlinegraphic{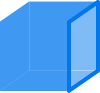}\label{eqn:twisted-stabilizers}
\end{align}
for a vertex $v$ and a cube $c$ on the intertwined boundary.

The trivial and intertwined boundaries are related to each other by $\U$~\eqref{eqn:UCX}, in the sense that $K_e \leftrightarrow X_e$ under the symmetry. Thus, applying $\U$ to the entire lattice simply swaps the two boundaries.

With these boundary conditions, the ITC encodes $K=2$ logical qubits. The counting of both the check operators and the constraints is easiest on a lattice with length $L$ in the periodic directions and length $L+1$ in the other direction. 
There are $3L^3+2L^2$ qubits on edges and $3L^3-L^2$ qubits on faces, for a total of $N = 6L^3 + L^2$ physical qubits. There are $L^2$ of the $K_e$ operators on the intertwined boundary, leaving $3L^3 + L^2$ edges to host $X_e$ operators, while all $3L^3-L^2$ faces have $Z_f$ operators. We find $L^3+L^2-1$ independent $A_v$ operators and $L^3-1$ independent $A_c$ operators. There are $3L^3+L^2$ total $B_f$ operators and $L^3+1$ relations between them, giving $2L^3+L^2-1$ independent $B_f$ operators. There are $3L^3 - 2L^2$ total $B_e$ operators and $L^3-L^2+1$ relations between them, leaving $2L^3-L^2-1$ independent $B_e$ operators. Thus, there are
\begin{align}
K = N - \frac{1}{2} \left( \log_2 |\G| + \log_2 |\S| \right) = 2
\end{align}
logical qubits. In the next subsection we construct bare and dressed logical operators for one of the logical qubits.

\subsection{Logical operators} \label{sub:logicals}

Recall that bare logical operators commute with all check operators (and therefore all stabilizers), while dressed logical operators commute with all stabilizers but not all check operators. 
In the bulk, we can construct membrane operators, 
\begin{align}
\barelogical{Z} = \prod_{f \in \cal{M}} Z_f, \qquad \barelogical{X} = \prod_{e \in \cal{M}^*} X_e, \label{eqn:logicals}
\end{align}
where $\cal{M}$ and $\cal{M}^*$ are a membrane and \emph{dual} membrane, respectively. If these membranes are contractible, then the associated operators are products of local stabilizers and therefore trivial logical operators. If the membranes are closed but noncontractible, then the operators are (possibly nonlocal) stabilizers, as previously discussed. To define logical operators, the membranes must terminate on open boundaries.

At the trivial boundary, all bare logical operators may terminate without violating any check operators. At the intertwined boundary, the naive termination of $\barelogical{X}$ commutes with all check operators, resulting in 
\begin{align}
\barelogical{X}=\inlinegraphic{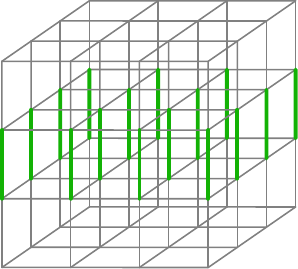}, \label{eqn:Xlog}
\end{align}
where, recall, the lattice is periodic in the top/bottom direction and the front/back direction. This operator is \emph{not} a product of check operators (and therefore not a stabilizer) because the single-site $X_e$ operators on the intertwined boundary is not in the check group $\G_\ITC$~\eqref{eqn:twisted-checks}.
However, the naive termination of $\barelogical{Z}$ anticommutes with the check operators $K_e$ for every edge $e$ in $\cal{C} = \partial \cal{M}\cap \text{Int}$, the intersection of the membrane boundary with the intertwined boundary. We can fix this by dressing the $\barelogical{Z}$ membrane operator with $\prod_{e\in \cal{C}} Z_e$, as in
\begin{align}
\barelogical{Z} = \inlinegraphic{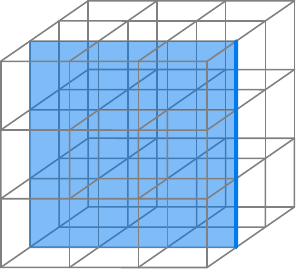}, \label{eqn:Zlog}
\end{align} 
which satisfies all the $K_e$ check operators. It is straightforward to verify that $\barelogical{X}$ and $\barelogical{Z}$ anticommute, showing that they are the bare logical operators for one of the two logical qubits.

The logical operators for the other logical qubit locally look like $\barelogical{X}$ and $\barelogical{Z}$, but are stretched in the opposite directions. If we name the new logical operators $\barelogical{X}_2$ and $\barelogical{Z}_2$, then $\barelogical{X}_2$ is now periodic in the top/bottom direction and $\barelogical{Z}_2$ is periodic in the front/back direction. This confirms that the ITC with periodic boundary conditions in the top/bottom direction and the front/back direction encodes two logical qubits. In the \hyperref[sec:appendix]{Appendix} we define the $e$-condensed ($m$-condensed) boundary, where only the logical $Z$ operator (logical $X$ operator) can terminate. We use these to define the ITC with only a single logical qubit.

The dressed logical operators are stringlike operators $\dressedlogical{Z} = \prod_{e \in \cal{C}} Z_e$ and $\dressedlogical{X} = \prod_{f \in \cal{C}^*} X_f$, where $\cal{C}$ and $\cal{C}^*$ are a noncontractible curve in the top/bottom direction and a noncontractible \emph{dual} curve in the front/back direction, respectively. These operators anticommute with the bare logical operators $\widebar{X}$ and $\barelogical{Z}$, but commute with each other. This is possible because they act on the logical qubits \emph{and} the nonlocal ``check qubits," as described previously.

In fact, the dressed logical operators are so named because they can be written as the bare logical operators dressed with check operators. For example the $Z_f$ operators in~\ref{eqn:Zlog} are all check operators. Multiplying $\barelogical{Z}$ by the individual $Z_f$ operators removes them, leaving behind just the line of $Z_e$ operators on the intertwined boundary. This is a dressed logical operator $\dressedlogical{Z}$. We can obtain other dressed logical operators by multiplying $\dressedlogical{Z}$ by various $B_f$ check operators, moving it into the bulk.

Whereas one of the logical operators in the 3d stabilizer toric code is inherently membranelike, the ITC possesses stringlike (dressed) logical operators for both sectors. Thus, we have traded some stability in one sector for extra stability in the other. In the next section, we will show how this stability results in detectability of partial logical operators.

\section{Errors and error correction} \label{sec:errors}

Having defined the check operators, stabilizers, and logical operators, we are prepared to describe the results of both qubit errors and measurement errors. We focus on $Z$-type qubit errors and $X$-type measurement errors, as the description of the other types follows analogously. One option is to describe the measurement outcomes of the $B$ operators and single-site check operators defined previously. This would result in measurement outcomes like those in Fig.~\ref{subfig:syndrome-toric}, with green for violated $X_e$ and pink for violated $B_e$. Instead, here we use a basis of $\G_\ITC$ that more closely mirrors the KVC, allowing us to use their decoder. We are only going to choose a different set of generators for the check group, leaving the check group itself unchanged, along with the lattice, geometric boundary conditions, stabilizers, and logical operators. Thus, we really are analyzing the same ITC, just with a different presentation. 

\begin{figure}
    \centering
    \includegraphics[width=.8\linewidth]{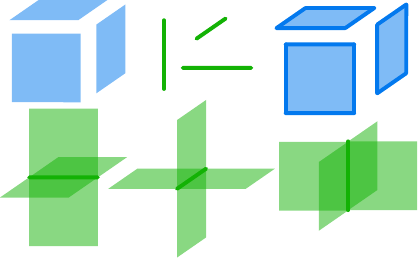}
    \caption{Check operators in an alternative presentation of the ITC. \textcolor{mygreen}{Green} and \textcolor{myblue}{blue} represent $X$ and $Z$ operators respectively. Solid bars are edges, and shaded parallelograms are faces. On the top row we have the operators $Z_f$, $X_e$, and $K_f$, while on the bottom row we have the $K_e$ check operators.}
    \label{fig:checks}
\end{figure}

On the same lattice, with qubits on edges and faces, define the operators 
\begin{align}
K_e &= X_e B_e = X_e \prod_{f\in \pardag e} X_f, \\
K_f &= Z_f B_f = Z_f \prod_{e \in \partial f} Z_e,
\end{align}
shown in Fig.~\ref{fig:checks}.
The check group
\begin{align}
\mathcal{G}_\ITC &= \left\langle X_e, Z_f,  K_e, K_f \right\rangle, \label{eqn:checks-color}
\end{align}
is the same as $\G_\ITC$ in~\eqref{eqn:checks-toric}.
The stabilizers are still the vertex and cube stabilizer operators,
\begin{align}
A_v &= \prod_{e \in \pardag v} X_e = \prod_{e \in \pardag v} K_e, \label{eqn:relation-A} \\
A_c &= \prod_{f \in \partial c} Z_f = \prod_{f \in \partial c} K_f,  \label{eqn:relation-B}
\end{align}
from~\eqref{eqn:stabilizers}, as they had to be.
The relations~\eqref{eqn:relation-A} and~\eqref{eqn:relation-B} between check operators provides redundancy that we lacked in the previous section. They allows us to detect measurement errors in addition to qubit errors, which improves the decoding procedure.

On the trivial boundary we have the same check operators as before. For a boundary face, the $K_f$ operator truncates to a four-body operator equivalent to $B_f$. For a boundary edge, naive truncation of $K_e$ leads to two body operators $X_e X_{f(e)}$ which we choose to discard. We instead keep $X_e$ for each boundary edge, resulting in the same set of check operators as before and the same stabilizers as in~\eqref{eqn:triv-stabilizers}.

On the intertwined boundary, we make the other choice and keep the two-body operator $K_e = X_e X_{f(e)}$, justifying the name we previously gave this operator. Thus, we arrive at the same check operators and stabilizers as previously in~\eqref{eqn:twisted-checks} and~\eqref{eqn:twisted-stabilizers}. 

In this section we describe the measurement outcomes of the KVC-inspired basis of measurements for the ITC. In Sec.~\ref{sub:local} we explore the ideal measurement outcomes that result from applying local qubit errors, using linear maps defined in Ref.~\cite{KubicaVasmer}. Then we apply some logical operators transversally (as a series of local errors) in Sec.~\ref{sub:transversal} and see which check operators they violate along the way. In Sec.~\ref{sub:single-shot} we outline how to use the measurement outcomes to perform single-shot error correction, even in the presence of measurement errors.

\subsection{Local errors} \label{sub:local}

Recall that any code state satisfies all stabilizers but violates many check operators. To construct a generic code state, start in a state $|X_0\rangle$ that satisfies all stabilizers and all $X$-type check operators. This state exists, as all these operators commute. Then, create a generic code state by acting with $Z$-type check operators on $|X_0\rangle$. The resulting state is still a code state and satisfies all stabilizers, but violates some $X$-type check operators. 
More specifically, consider a $Z$-type check operator $\eta$ and the linear map $\delta_M$, which maps $\eta$ to the set of check operators that anticommute with it. The nontrivial measurement outcome of $X$-type measurements on the state $\eta |X_0\rangle$ is $\delta_M \eta$.

To illustrate this, let the $Z$-type check operator be $\eta = K_f$ so that the measurement outcome is $\delta_M\eta = \prod_{e\in \partial f} X_e$, four $X_e$ operators around the face $f$. Graphically, this relationship is
\begin{align}
\vcenter{\hbox{\includegraphics[scale=.9]{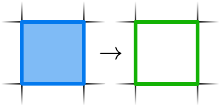}}}, \label{eqn:bulk-error-A}
\end{align}
with $\eta$ on the left and $\delta_M \eta$ on the right. Similarly, the $Z_f$ check operator anticommutes with the four $K_e$ operators on edges around the face $f$. This relationship is
\begin{align}
\vcenter{\hbox{\includegraphics[scale=.9]{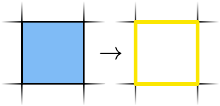}}}, \label{eqn:bulk-error-B}
\end{align}
where \textcolor{myyellow}{yellow} lines are violated $K_e$ check operators. By stacking many check operators we can create the measurement outcome of a generic code state, which consists of many green and yellow lines, all of which must form closed loops. 

For a vertex $v$ on the trivial boundary (chosen to be on the left hand side) we have to measure the additional operator
\begin{align}
A_v^\text{2d, eff} = \inlinegraphic{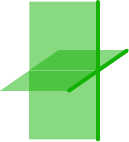}, \label{eqn:twist-RBH}
\end{align}
which can be shown to be in the check group. Measuring this operator allows for two redundant ways to construct the $A_v$ operator out of measurement outcomes for $v$ on the trivial boundary.\footnote{Thank you to Yaodong Li for pointing out the necessity of measuring this operator.} In particular, the redundancy is
\begin{align}
A_v = \prod_{e\in \pardag v} X_e = A_v^\text{2d, eff} K_{e(v)},
\end{align}
where $e(v)$ is the unique nonboundary edge in $\pardag v$. Let us draw violated $A_v^\text{2d, eff}$ operators as short yellow lines sticking out of the lattice.
Green lines still cannot end, but we can see from 
\begin{align}
\vcenter{\hbox{\includegraphics[scale=.9]{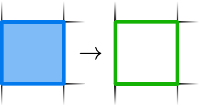}}}, \quad \vcenter{\hbox{\includegraphics[scale=.9]{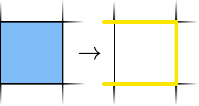}}} 
\end{align}
that yellow lines can end if they extend to the left of the boundary. 

On the intertwined boundary (on the right side) we must measure the redundant operator
\begin{align}
A_v^\text{2d} = \inlinegraphic{triv-Para},
\end{align}
analogous to~\ref{eqn:twist-RBH}. The redundancy in stabilizer operators is
\begin{align}
A_v = \prod_{e \in \pardag v} K_e = A_v^\text{2d} X_{e(v)},
\end{align}
for a vertex $v$ on the intertwined boundary.
Let us draw violations of $A_v^\text{2d}$ as green lines extending beyond the boundary.
Measurement outcomes such as
\begin{align}
\vcenter{\hbox{\includegraphics[scale=.9]{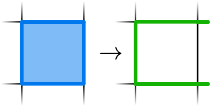}}}, \quad \vcenter{\hbox{\includegraphics[scale=.9]{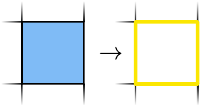}}}
\end{align}
tell us that now green lines can end if they extend to the right of the boundary. To summarize, the violated check operators in any code state consist of closed green ($X_e$) and yellow ($K_e$) loops in the bulk. Yellow loops may end at the trivial boundary and green lines may end at the intertwined boundary. 

To leave the code space, act with an operator that anticommutes with some stabilizers. Any such operator, called an error, is not in the check group. We can extend the map $\delta_M$ to map an error $\epsilon$ to the check operators $\delta_M\epsilon$ that it anticommutes with. Another map, called $\partial_S$, maps an error to a stabilizer syndrome, the set of stabilizers it anticommutes with. We can represent both maps graphically, simultaneously. For example, in
\begin{align}
\vcenter{\hbox{\includegraphics[scale=.9]{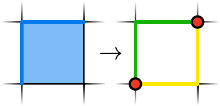}}},
\end{align}
we have $\epsilon$ on the left and both $\delta_M \epsilon$ and $\partial_S \epsilon$ on the right.
The $Z$-type error on the left anticommutes with two $X_e$ check operators (\textcolor{mygreen}{green}), two $K_e$ check operators (\textcolor{myyellow}{yellow}), and two $A_v$ stabilizers (\textcolor{myred}{red}). 
A different error,
\begin{align}
\vcenter{\hbox{\includegraphics[scale=.9]{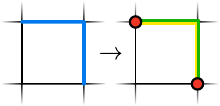}}}
\end{align}
also maps to two $X_e$ check operators, two $K_e$ check operators, and two $A_v$ stabilizers. 

Composing similar errors creates generic syndromes and measurement outcomes, subject to some constraints called relations. The relations say that green lines and yellow lines can only end on red points, and red points must have an odd number of green lines and an odd number of yellow lines incident upon them. Phrased in terms of operators, the relation is that, for any vertex $v$, the number of violated $K_e$ and $X_e$ operators on the edges in $\pardag v$ must have the same parity. If that parity is odd then the stabilizer $A_v$ is violated, and if the parity is even then $A_v$ is satisfied. These relations follow from~\eqref{eqn:relation-A} and~\eqref{eqn:relation-B}. In other words, violated stabilizers are connected by violated check operators, so that they are no longer purely pointlike errors. In Sec.~\ref{sub:single-shot} we show how these connecting lines allow us to probabilistically determine the locations of the true qubit errors, even in the presence of measurement errors. 

On the trivial boundary we know that yellow lines can end without violating stabilizers. Errors cause measurement outcomes that look like
\begin{align}
\vcenter{\hbox{\includegraphics[scale=.9]{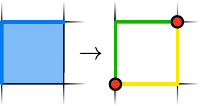}}}, \quad \vcenter{\hbox{\includegraphics[scale=.9]{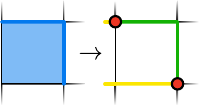}}},
\end{align}
which show that violated stabilizers still need green and yellow lines attached to them, but yellow lines may end without a corresponding violated stabilizer to the left of the boundary.
The intertwined boundary instead allows configurations like
\begin{align}
\vcenter{\hbox{\includegraphics[scale=.9]{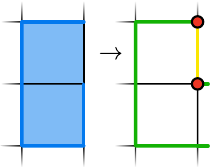}}}, \quad \vcenter{\hbox{\includegraphics[scale=.9]{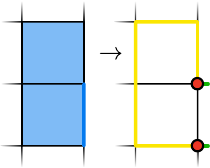}}},
\end{align}
so that green lines can end without a corresponding violated stabilizer.
To summarize, red dots (violated $A_v$ syndromes) need to be connected to green and yellow lines. Yellow lines may end without a violated stabilizer at the trivial boundary and green lines may end without a violated stabilizer at the intertwined boundary. 

In the coarse-grained description of the ITC, any $Z$-type error consists of open strings of $Z_e$ operators, possibly decorated by open or closed membranes of $Z_f$ operators and closed strings of $Z_e$ operators. A bare string of $Z_e$ operators anticommutes with $X_e$ and $K_e$ check operators along its length. A bare membrane of $Z_f$ operators anticommutes with $K_e$ check operators around its boundary, and a $Z_f$ membrane dressed with $Z_e$ operators around its boundary anticommutes with $X_e$ check operators around that boundary. Open $Z_e$ strings violate $A_v$ stabilizers at their endpoints. These rules allow us to invert the $\delta_M$ map and find a representative error that can cause any valid measurement outcome. For example, the measurement outcome represented graphically in Fig.~\ref{subfig:syndrome-color} could be caused by the error shown in Fig.~\ref{subfig:syndrome-error}.

\subsection{Transversal logical errors} \label{sub:transversal}

Isolated errors are not so dangerous. Problems arise when errors conspire to mimic logical operators. For example, applying a string of $Z_e$ errors is the same as applying a partial $Z$-type dressed logical operator. Applying a logical operator this way, as a series of local operators, is called a transversal decomposition of the logical operator. At any point in the decomposition, the partial logical operator results in violated $A_v$ stabilizers at its endpoints. Thus, a transversal decomposition of the $Z$-type dressed logical operator transports violated $A_v$ operators across the system, leaving behind violated $X_e$ and $K_e$ check operators. This subsection is a pedagogical description of the transversal decomposition of the bare logical $Z$ operator.

Recall the $Z$-type bare logical operator
\begin{align}
\inlinegraphic{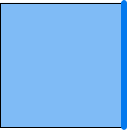},
\end{align}
drawn as a slice through the entire lattice.
The figure represents a coarse-grained version of the one in~\eqref{eqn:Zlog}. We could apply this operator transversally in many different ways. We could truncate it from right to left,
\begin{align}
\inlinegraphic{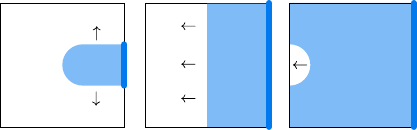}, \label{eqn:trans-log-A}
\end{align}
resulting in the measurement outcomes
\begin{align}
\inlinegraphic{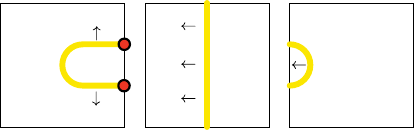},
\end{align}
where the noncontractible strings of violated $K_e$ check operators witnesses the partial logical operator. A geometrically different transversal decomposition of the bare logical operator,
\begin{align}
\inlinegraphic{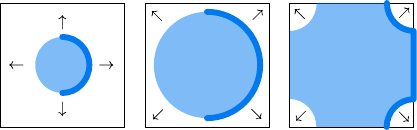}, \label{eqn:trans-log-B}
\end{align}
results in the measurement outcomes 
\begin{align}
\inlinegraphic{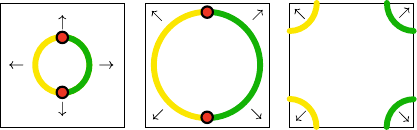},
\end{align}
where there are noncontractible strings of violated $K_e$ and $X_e$ check operators stretching from one boundary to the other. A different transversal decomposition of $\barelogical{Z}$ uses a different boundary,
\begin{align}
\inlinegraphic{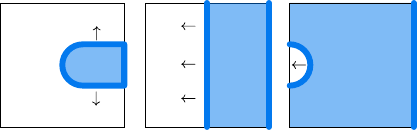}, \label{eqn:trans-log-C}
\end{align}
resulting in the measurement outcomes
\begin{align}
\inlinegraphic{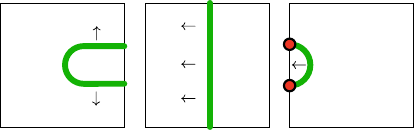},
\end{align}
where the noncontractible strings now consist of violated $X_e$ checks. Note that the violated stabilizers can appear on the left or right boundary or in the bulk, depending on the decomposition.

What is constant is that all decompositions involve transporting a pair of violated stabilizers vertically across the system and then removing the resulting violated check operators at the left and right boundary. Thus, either type of logical operator transports violated stabilizer across the system, and the difference is that dressed logical operators leave violated check operators behind while bare logical operators remove the violated check operators at the boundary.

\subsection{Single-shot error correction} \label{sub:single-shot}

The point of any error-correction procedure is to prevent local errors from accumulating into logical errors. To that end, we want to detect and undo any partial logical operators. The simplest way is to find all violated stabilizers and pair them up to correct them while introducing the fewest residual errors. One way to do this in the stabilizer toric code is the minimal-weight perfect matching (MWPM) procedure~\cite{DennisTopological}. 

The MWPM procedure successfully decodes the stabilizer toric code as long as measurements are ideal. However, incorrect measurements, where the wrong measurement outcome is written down, completely ruin the MWPM procedure for pointlike stabilizers. The problem is that a small number of incorrect measurements can lead to a large number of residual errors. 

\begin{figure}
    \centering
    \includegraphics[width=\linewidth]{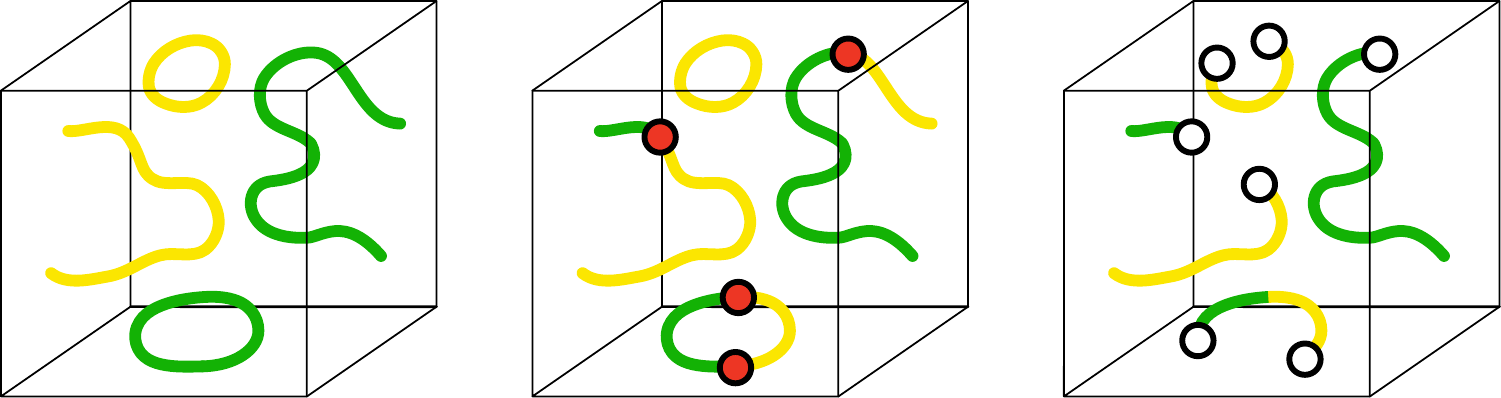}
    \caption{Syndromes in the KVC-inspired presentation of the ITC. On the left we have a syndrome for a code states with no qubit or measurement errors. The middle figure displays qubit errors, which are violated stabilizers represented by red dots, but no measurement errors. On the right we have measurement errors, which are violated relations represented by empty circles. }
    \label{fig:states}
\end{figure}

Connecting violated stabilizers with violated check operators, as in the GCC, the KVC, and the ITC, helps to recover from measurement errors. In fact, the syndromes of the ITC look like the syndromes of the KVC, at least when coarse-grained as in Fig.~\ref{fig:states}. This allows us to use the single-shot version of the MWPM procedure introduced in Ref.~\cite{KubicaVasmer}. 

The single-shot MWPM procedure consists of two steps, which we describe first heuristically and then in more mathematical detail. The first step deals with violations of the relations from Sec.~\ref{sub:local}, which are unpaired endpoints of the measurement outcomes. These violations, collectively called the relation syndrome, are aphysical and can only result from measurement error. For example, if the true measurement outcome is a closed line of violated $X_e$ check operators, some false negatives along this line result in endpoints of the line. The relation syndromes are the white circles in Fig.~\ref{fig:states}. The first round of MWPM pairs up these violated relations, inferring the shortest strings of measurement errors that could have caused them. 

After inferring the measurement errors, we are left with valid measurement outcomes, with their attendant stabilizer syndromes. Recall that violated stabilizers are vertices where $X_e$ and $K_e$ lines simultaneously end. These are marked with red dots in Fig.~\ref{fig:states}. The second round of MWPM pairs up the violated stabilizers, matching the stabilizers in a way that is robust to measurement errors. Finally, error correction proceeds by acting with operators that remove the violated stabilizers.

To describe the process in more detail, the next part of this section unifies the entire error-correction procedure as a series of linear maps between vector spaces, closely following the discussion of single-shot error correction in Ref.~\cite{KubicaVasmer}. We outline the important steps, highlighting the places where the ITC differs from the KVC.

We have been describing the ITC in terms of sets and groups: the set of qubits, the check group, the stabilizer group, the set of relations, etc. To formalize the decoding problem, we redefine these as vector spaces over $\mathbb{F}_2$, the field with two elements. These spaces are the space of $Z$-type check operators $C_G$, the space of qubits $C_Q$, the space of $X$-type check operators $C_M$, the space of stabilizers $C_S$, and the space of relations $C_R$.
For example, an element of $C_S$ corresponds to an element of the stabilizer group, and an element of $C_Q$ corresponds to a subset of the qubits. An ``error" involves a $Z$-type qubit error $\epsilon \in C_Q$ and a measurement error $\mu \in C_M$, while a ``syndrome" includes both the violated stabilizers $\sigma \in C_S$ and the violated relations in $\omega \in C_R$.

We already defined some linear maps between these spaces in Sec.~\ref{sub:local}: $\delta_M$ maps an error $\epsilon \in \C_Q$ to the set of check operators $\delta_M\epsilon \in \C_M$ that it anticommutes with, $\delta_S$ maps a measurement outcome $\zeta \in \C_M$ to the inferred syndrome $\delta_S \zeta \in C_S$ and $\partial_S$ maps an error $\epsilon$ to the syndrome $\partial_S \epsilon\in C_S$ that it causes. These maps obey 
\begin{align}
\partial_S = \delta_S \delta_R
\end{align}
by definition.
We also define some new maps: $\partial_Q$ maps a $Z$-type check operator $\gamma \in C_G$ to the set of qubits $\partial_Q\gamma \in C_Q$ on which it acts, and $\delta_R$ maps a measurement outcome $\zeta$ to its relation syndrome $\delta_R \zeta \in C_R$.

The above-defined maps must obey some constraints coming from the definition of subsystem codes. First, the constraint 
\begin{align}
\partial_S \partial_Q = 0
\end{align}
says that the $Z$-type check operators commute with the $X$-type stabilizers. Second, the constraint 
\begin{align}
\delta_R \delta_M = 0
\end{align}
says that physical errors do not cause relation syndromes. These constraints are all encoded in a commutative diagram taken directly from Ref.~\cite{KubicaVasmer},
\begin{widetext}
\begin{align}
\begin{tikzcd}[ampersand replacement=\&, column sep = 0, row sep = 0]
\&[6em] \& \begin{array}{c} \text{$X$-type check} \\ \text{measurements} \end{array}  \& \\
\&\& C_M \ar[dr, "{\left(\begin{array}{c} \delta_S\\\hline \delta_R\end{array}\right)}"] \&\\[10ex] 
C_G \ar[r, "{\left(\begin{array}{c} \partial_Q\\\hline 0\end{array}\right)}"] \&
C_Q \oplus C_M \ar[ur, "{\left(\begin{array}{c|c} \delta_M &I\end{array}\right)}"]
\ar[rr, "{\left(\begin{array}{c|c} \partial_S& \delta_S \\\hline 0& \delta_R \end{array}\right)}"]\&\& C_S \oplus C_R\\
\begin{array}{c} \text{$Z$-type} \\ \text{check operators} \end{array} \&
\begin{array}{c} \text{qubits and $X$-type} \\ \text{measurements} \end{array} \& \&
\begin{array}{c} \text{stabilizers} \\ \text{and relations} \end{array}
\end{tikzcd}  \label{eqn:cd}
\end{align}
\end{widetext}
where the block matrix structure of the maps follows from treating elements of the vector spaces as column vectors.

In the ITC, these spaces and maps decompose even further, simplifying the discussion. For example, the space of $Z$-type check operators is $C_G=C_{Zf}\oplus C_{Kf}$, the direct sum of the spaces of $Z_f$ and $K_f$ check operators. The space of qubits is $C_Q=C_E \oplus C_F$, the direct sum of the spaces of edge qubits and face qubits, with $C_{Zf}$, $C_{Kf}$, and $C_F$ all isomorphic to the space of faces in the lattice. The space of $X$-type check operators decomposes into $C_M = C_{Xe} \oplus C_{Ke}$, the direct sum of the spaces of $X_e$ and $K_e$ check operators, with  $C_{Xe}$, $C_{Ke}$, and $C_E$ all isomorphic to the space of edges on the lattice. The space of stabilizers $C_S$ and the space of relations $C_R$ are both isomorphic to the space of vertices.

Each block of the block matrices in~\eqref{eqn:cd} alse decomposes into a block structure. Furthermore, the individual components are simply the identity $I$ and the geometric boundary operator $\partial$ in the bulk, although these may have to be truncated or supplemented at either boundary. The first column of\begin{align}
\partial_Q = \left( \begin{array}{c|c} 0 & \partial \\\hline I & I
\end{array} \right)
\end{align}
says that the $\Z_f$ check operator acts on a single face qubit, while the second column says that $K_f$ acts on a single face qubit and the four edge qubits that surround it. Similarly, the first column of
\begin{align}
\delta_M = \left( \begin{array}{c|c} I & 0 \\\hline I & \partial \label{eqn:delta-M}
\end{array} \right)
\end{align}
says that a $Z_e$ error anticommutes with the $X_e$ and $K_e$ check operator on that edge, while the second column says that a $Z_f$ error anticommutes with the four $K_e$ check operators around that face. Equations~\eqref{eqn:bulk-error-A} and~\eqref{eqn:bulk-error-B} represent~\eqref{eqn:delta-M} graphically.
The two maps 
\begin{align}
\delta_S = \left( \begin{array}{c|c} \partial & 0
\end{array} \right), \qquad \delta_R = \left( \begin{array}{c|c} \partial & \partial
\end{array} \right)
\end{align}
say that violated stabilizers are vertices with an odd number of violated $X_e$ check operators around them, while violated relations are vertices with an odd total number of violated $X_e$ and $K_e$ check operators around them, respectively.
The last map, $\partial_S = \left( \begin{array}{c|c} \partial & 0
\end{array} \right)$, maps a set of qubit errors directly to the stabilizers it violates. 

The decompositions automatically satisfy the constraints $\partial_S = \delta_S \delta_M$, $\partial_S \partial_Q=0$, and $\delta_R\delta_M = 0$. The point of including this whole decomposition is to show that the decoding maps in the ITC are nicely geometric, like the decoding maps in the stabilizer toric codes.

The decoding strategy takes two steps. As described previously, code states have nontrivial syndromes generated by $\gamma \in C_G$, a set of $Z$-type check operators. The initial $\gamma$ and some errors $\epsilon \oplus \mu \in C_Q \oplus C_M$ produce a 
measurement outcome 
\begin{align}
\zeta = \delta_M \epsilon + \mu + \delta_M\partial_Q \gamma \in C_M.
\end{align}
This measurement outcome violates a set of relations $\delta_R \zeta$, allowing us to infer a measurement error $\hat{\mu}$ with the same relation syndrome $\delta_R \hat{\mu} = \delta_R \zeta$. The resulting inferred stabilizer syndrome is $\delta_S (\zeta + \hat\mu)$, allowing us to infer a qubit error $\hat{\epsilon}$ such that $\partial_S \hat{\epsilon} = \delta_S (\zeta + \hat\mu)$. After correcting for the inferred error, the residual error is $\epsilon + \hat \epsilon$. If the decoding procedure succeeds perfectly, we have $\hat\epsilon = \epsilon + \partial_Q \gamma'$, where $\gamma' \in C_G$, so that the residual stabilizer syndrome is $\partial_S(\epsilon + \hat \epsilon) = \partial_S \partial_Q \gamma' = 0$. This is the doubled MWPM procedure from Ref.~\cite{KubicaVasmer}. Even if the decoding procedure does not proceed perfectly, the residual stabilizer syndrome should be equivalent to one caused by a small number of qubit errors.

An advantage of the ITC is that both rounds of MWPM occur on the graph defined by the edges and vertices of the cubic lattice. In both cases, the violations appear on vertices and the inferred errors appear on edges. Similarly, for $X$-type qubit errors and $Z$-type measurement errors, the violations appear on cubes and the inferred errors appear on faces.
This is geometrically dual to the previous cases. In all cases, the problems of finding $\hat \mu$ and $\hat \epsilon$ simply reduce to MWPM on the cubic lattice. At this point we can rely on the results from Ref.~\cite{KubicaVasmer} to show that the ITC is indeed capable of single-shot error correction.

\begin{figure}[t]
    \centering
    \subfigure[\label{subfig:syndrome-error}]{\includegraphics[width=.49\linewidth]{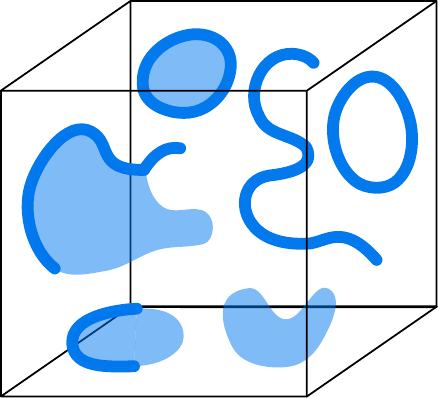}}
    \hfill
    \subfigure[\label{subfig:syndrome-overcomplete}]{\includegraphics[width=.49\linewidth]{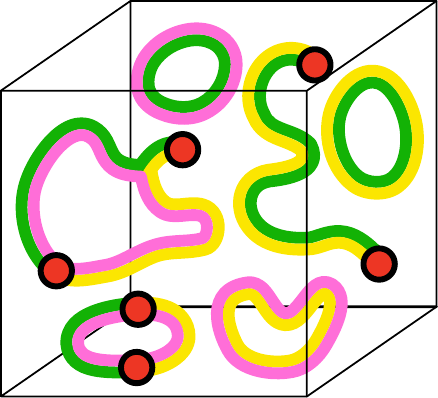}}
    \caption{An error in the ITC and its overcomplete syndrome. (a) If shaded blue areas are $Z_f$ operators and blue lines are $Z_e$ operators, this is the error causing the syndromes shown in Fig.~\ref{fig:syndrome}. (b) Measuring an overcomplete set of check operators leads to maximally redundant syndromes, where every line must be coincident with a line of another color. Note that removing yellow lines gives Fig.~\ref{subfig:syndrome-toric} while removing pink lines gives Fig.~\ref{subfig:syndrome-color}.}
    \label{fig:syndrome-overcomplete}
\end{figure}

Moving beyond the analysis of Ref.~\cite{KubicaVasmer},
we can achieve maximal redundancy by measuring all six types of check operators: $X_e$, $Z_f$, $K_e$, $K_f$, $B_e$, and $B_f$. This set is overcomplete in the sense that to every edge there corresponds three check operators, with a nontrivial relation between them: $K_e X_e B_e = 1$. We can use the previous analysis to infer the relations for $X$-type check operators in the bulk. Every edge must be occupied by zero or two violated check operators. Lines of violated $B_e$ check operators may not end, while $K_e$ and $X_e$ lines only end on violated stabilizers. Violated stabilizers must be simultaneous endpoints of $K_e$ and $X_e$ lines. These rules are illustrated in Fig.~\ref{fig:syndrome-overcomplete}. On the trivial boundary, $K_e$ and $B_e$ lines may end without violated stabilizers, and violated stabilizers only need to be endpoints of $X_e$ lines. On the intertwined boundary, $X_e$ and $B_e$ lines may end wihtout violated stabilizers, and violated stabilizers only need to be endpoints of $K_e$ lines.

In this overcomplete version of the ITC, the nontrivial relation $K_e X_e B_e = 1$ for every edge $e$ means that a string of measurement errors is detectable everywhere along its length, rather than just at its endpoints. This could possibly lead to a more accurate while still efficient decoder. Of course, measuring overcomplete sets of check operators can always increase the accuracy of a decoder. This realization is not new, but might help in understanding the ITC.

\section{Phase Diagram of the ITC} \label{sec:phase}

So far, we have seen that the check group $\G_\ITC$ includes some terms that we might recognize from the our favorite Hamiltonians. By construction, the $B$ terms come from the 3d toric code Hamiltonian~\cite{CastelnovoChamon}. It is easy to see that the single-site terms form a paramagnet Hamiltonian. In addition, the $K$ terms are equivalent to terms in the Raussendorf-Bravyi-Harrington (RBH) Hamiltonian~\cite{RaussendorfLongRange} after a Hadamard transformation on face qubits, mapping $X_f \leftrightarrow Z_f$. In this section we show that these three Hamiltonians live in the phase diagram corresponding to the ITC subsystem code.

\subsection{Gapped phases in the ITC} \label{sub:phases}

Recall that, from a stabilizer code with stabilizer group $\S$, we can construct a gapped Hamiltonian,
\begin{align}
H_\S = -\sum_{S \in \S} J_S S, \label{eqn:H-S}
\end{align}
by including all the stabilizers with negative coefficients $-J_S < 0$. The coefficients for operators with large support should be small.
Because the stabilizers all commute, the relative strengths of the coefficients do not matter. Thus, the stabilizer code corresponds to a single phase of matter. We can try to build a Hamiltonian from a subsystem code analogously, by including all the stabilizers and check operators with coefficients. The most general such Hamiltonian is
\begin{align}
H_\G = -\sum_{S \in \S} J_S S - \sum_{G \in \G} J_G G + \text{h.c.}, \label{eqn:H-G}
\end{align}
where we should again be careful to bound the strength of the coefficients for operators with large support~\cite{EllisonSubsystem}. The stabilizer coefficients $J_S$ should be negative, but the check group contains pure phases so there is no reason to impose negativity on the check coefficients $J_G$. 

Since the check group is non-Abelian, the Hamiltonian $H_\G$ is frustrated and different choices of coefficients may give rise to different gapped phases. Thus, the subsystem code corresponds to a whole phase diagram with different phases and phase transitions. Phase diagrams have been studied for the GCC~\cite{KubicaYoshida} and the KVC~\cite{LiPhaseDiagram}. Additionally, subsystem codes can lead to novel constructions for 2d Abelian phases~\cite{EllisonSubsystem}.

To simplify the analysis of the ITC Hamiltonian, define the Hamiltonian
\begin{align}
H_\text{ITC} = &- J_{Av} \sum_v A_v - J_{Ac} \sum_c A_c \nonumber\\
& - J_{Xe} \sum_e X_e - J_{Ke} \sum_e K_e - J_{Be} \sum_e B_e \nonumber\\
& - J_{Zf} \sum_f Z_f - J_{Kf} \sum_f K_f - J_{Bf} \sum_f B_f,
\end{align}
which has fewer tunable parameters. This simplification means that we might not find all of the phases in the full Hamiltonian~\eqref{eqn:H-G}, but we can still find several interesting phases. As the $A_v$ and $A_c$ terms commute with everything else, we can choose to set $J_{Av}=J_{Ac}=\infty$. This in turn is equivalent to enforcing a symmetry. In fact, $A_v$ and $A_c$ each generate a 1-form symmetry, so we realize we are studying 1-form symmetry-protected topological (SPT) phases. 

To find some phases of $H_\ITC$, tune to a limit where some $J_G$ are much larger than others. If the largest terms all commute, the resulting Hamiltonian is exactly solvable. For example, if the coefficients of the single-site terms $J_{Xe}$ and $J_{Zf}$ are much larger than the rest of the check coefficients, then the effective Hamiltonian is the paramagnet Hamiltonian
\begin{align}
H_\Para = - J_{Xe} \sum_e X_e - J_{Zf} \sum_f Z_f, \label{eqn:H-para}
\end{align}
with some small perturbations. Some other exactly-solvable Hamiltonians are
\begin{align}
H_{\TCE} &= - J_{Av} \sum_v A_v - J_{Zf} \sum_f Z_f - J_{Bf} \sum_f B_f, \label{eqn:H-TCE}\\ 
H_{\TCF} &= - J_{Ac} \sum_c A_c - J_{Xe} \sum_e X_e - J_{Be} \sum_e B_e, \label{eqn:H-TCF}
\end{align}
which each represent a single copy of the toric code stacked with a paramagnet, and
\begin{align}
H_\TTC =& - J_{Av} \sum_v A_v - J_{Ac} \sum_c A_c \nonumber\\
&- J_{Be} \sum_e B_e - J_{Bf} \sum_f B_f, \label{eqn:H-2TC}
\end{align}
which represents the two toric codes from~\eqref{eqn:S-2TC}. Some distinct choices result in the same phases. For example, the Hamiltonian
\begin{align}
H = - J_{Av} \sum_v A_v - J_{Zf} \sum_f Z_f - J_{Kf} \sum_f K_f,
\end{align}
belongs to the same phase as $H_\TCE$ because $Z_fK_f = B_f$. 

Another useful limit is the RBH Hamiltonian
\begin{align}
H_\RBH = - J_{Ke} \sum_e K_e - J_{Kf} \sum_f K_f, \label{eqn:H-RBH}
\end{align}
whose ground state is the 3d cluster state~\cite{RaussendorfLongRange}. This Hamiltonian is in the trivial phase, but in a nontrivial SPT phase under the 1-form symmetry generated by $A_v$ and $A_c$.
Altogether, we find five phases in $H_\ITC$, represented by the five named Hamiltonians above.

Some of the exactly-solvable Hamiltonians have previously been shown to lead to symmetry-protected self-correction. These are $H_\RBH$~\cite{RobertsBartlett}, $H_\Para$~\cite{StahlNandkishore}, and $H_\TTC$~\cite{StahlSymmetry}. In each case the protecting symmetry is (possibly a subset of) the 1-form symmetry generated by $A_v$ and $A_c$. However, in all three cases the boundary conditions must be chosen differently to lead to symmetry-protected self-correction rather than single-shot error correction. It is likely that the TCE and TCF phases would also support self-correction with the right choice of boundary and 1-form symmetry. In addition, the RBH phase supports symmetry-protected topological order at nonzero temperatures~\cite{RobertsSPTO}, and the other phases likely do as well.

\subsection{Topological order in each phase} \label{sub:topological-order}

Every phase that Ref.~\cite{LiPhaseDiagram} finds in the phase diagram of the KVC has topological order somewhere, either in the bulk or on the boundary. In the ITC, the phases corresponding to $H_\TCE$, $H_\TCF$, and $H_\TTC$ are topologically ordered in the bulk. In the paramagnetic and RBH phases, the topological order instead must occur on the boundary. The symmetry and the bulk Hamiltonian together restrict the possible gapped boundaries, as is common in SPTs. For example, consider the paramagnetic phase. On the trivial boundary, we can include an $X_e$ term for every boundary edge without causing problems, giving paramagnetic boundary conditions. On the intertwined boundary, however, such a term is not in $\G_\ITC$ and would anticommute with the stabilizers. Instead, we must include the terms 
\begin{align}
A_v^\text{2d} = \inlinegraphic{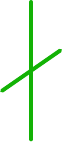}, \qquad B_f^\text{2d} = \inlinegraphic{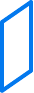},
\end{align}
which make up a 2d toric code on this boundary. We can check that these terms are in $\G_\ITC$.
Similarly, in the RBH phase, we can include $K_e$ for edges in the intertwined boundary, leading to no topological order there. The trivial boundary must instead have the terms
\begin{align}
A_v^\text{2d, eff} = \inlinegraphic{twist-RBH}, \qquad B_f^\text{2d} = \inlinegraphic{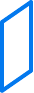},
\end{align}
which define an effective 2d toric code there. These terms are in $\G_\ITC$ and define the standard toric code boundary conditions of the RBH Hamiltonian.

In the TCE and 2TC phases the boundary conditions are just the ``smooth" boundaries of the 3d toric codes, where membrane operators may terminate but stringlike operators may not. In the TCF phase this is still the case, but it is less clear on the intertwined boundary. If we include the $K_e$ operators on the boundary edges, then the naive termination of $A_c$ does not commute with these. Instead, we use the dressed termination of $K_e$ from~\eqref{eqn:twisted-checks} and $A_c$ from~\eqref{eqn:twisted-stabilizers}, so that the terms in the Hamiltonian are
\begin{align}
\inlinegraphic{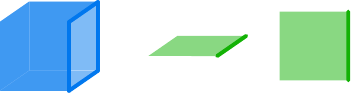}, \label{eqn:twisted-TCF}
\end{align}
along with the terms in the bulk and on the trivial boundary.
This new boundary looks funny, but is still a valid (smooth) termination of the 3d toric code on faces. It is related to the ordinary termination (on the trivial boundary) by the $\U$ symmetry~\eqref{eqn:UCX}.

Reference~\cite{LiPhaseDiagram} additionally shows that the bare logical operators in the subsystem code act as logical operators in every phase. In the ITC, this is particularly straightforward in the paramagnetic phase; $\barelogical{X}$~\eqref{eqn:Xlog} and $\barelogical{Z}$~\eqref{eqn:Zlog} consist of the boundary $X$-type logical operator dressed by $X_e$ terms and the boundary $Z$-type logical operator dressed by $Z_f$ terms, respectively. In the paramagnetic phase, the single-site $X_e$ and $Z_f$ terms are stabilizers, so $\barelogical{X}$ and $\barelogical{Z}$ remain logical operators.

In the RBH phase the relationship is not as obvious. To see that it still holds, consider a product of $X$-type $K_e$ stabilizers,
\begin{align}
\inlinegraphic{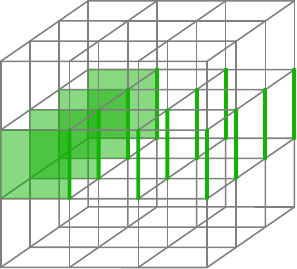},
\end{align}
and an $X$ logical operator on the boundary toric code,
\begin{align}
\inlinegraphic{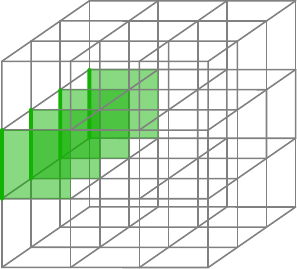},
\end{align}
recalling that both figures are meant to be periodic in the front/back and top/bottom directions. Multiplying these together does in fact give $\barelogical{X}$, as in~\eqref{eqn:Xlog}. Similarly a product of $Z$-type $K_f$ stabilizers,
\begin{align}
\inlinegraphic{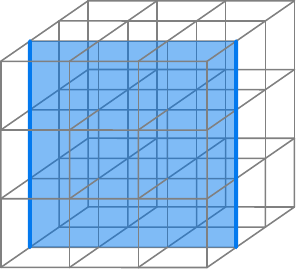},
\end{align}
and a $Z$ logical operator,
\begin{align}
\inlinegraphic{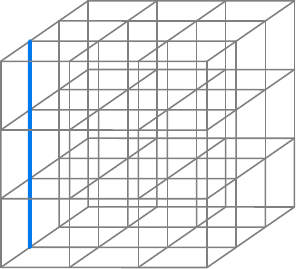},
\end{align}
multiply $\barelogical{Z}$, as in~\eqref{eqn:Zlog}.

In the TCE phase, $\barelogical{X}$ is the membrane logical operator and $\barelogical{Z}$ is the stringlike logical operator dressed with single-site $Z_f$ stabilizers. In the TCF phase, $\barelogical{Z}$ is the membrane logical operator, whose dressing is required as in~\eqref{eqn:twisted-TCF}. The single-site $X_e$ operators in $\barelogical{X}$ are all individual stabilizers except for those on the intertwined boundary. The latter can be multiplied by $K_e$ on every edge to give the stringlike logical operator in the TCF phase. 
Since the 2TC phase just consists of the toric code on edges stacked with the toric code on faces, it is clear that $\barelogical{X}$ acts only on one code while $\barelogical{Z}$ acts on both. Thus, one of the logical qubits in the 2TC phase is privileged over the other. It would be interesting to find a more systematic understanding of the relationship between logical qubits in the parent subsystem code and logical qubits in individual gapped phases.

\subsection{Constructing phase diagrams} \label{sub:phase-diagrams}

\begin{figure}
    \centering
    \includegraphics[width = \linewidth]{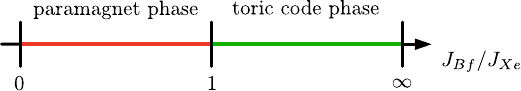}
    \caption{Phase diagram for the Hamiltonian in~\eqref{eqn:1d-phase}. Within the $A_v$- and $Z_f$-satisfying Hilbert space, the $B_f$ and $X_e$ terms are dual to each other, so that the phase transition occurs when the couple strengths are equal. The system enters the 3d toric code phase when $J_{Bf}$ is large and enters the paramagnetic phase when $J_{Xe}$ is larger.}
    \label{fig:1d-phase}
\end{figure}

So far, we have been analyzing individual phases via exactly-solvable Hamiltonians.
By including noncommuting terms with comparable strength, we can instead build Hamiltonians with phase transitions. For example, the Hamiltonian
\begin{align}
H = &- J_{Av} \sum_v A_v - J_{Zf} \sum_f Z_f \nonumber\\
& - J_{Bf} \sum_f B_f - J_{Xe} \sum_e X_e \label{eqn:1d-phase}
\end{align}
describes 3d $\mathbb{Z}_2$ lattice gauge theory; the first two terms commute with everything and are always satisfied, but the second two terms compete. Within the space of states satisfying the first two terms, the second two terms are dual~\cite{FradkinSusskind, WegnerDuality}, so that the phase transition happens at $J_{Bf} = J_{Xe}$~\cite{Creutz}.  When $J_{Bf}>J_{Xe}$ the Hamiltonian is in the TCE phase, while when $J_{Bf}<J_{Xe}$ the Hamiltonian is in the paramagnetic phase, as shown in Fig.~\ref{fig:1d-phase}. 

\begin{figure}
    \centering
    \includegraphics[width = \linewidth]{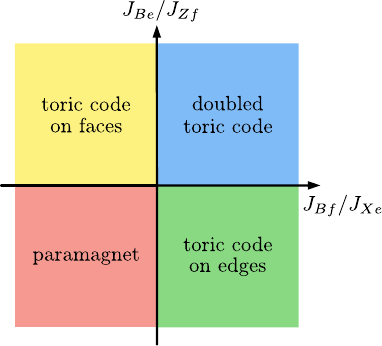}
    \caption{Phase diagram for the Hamiltonian in~\eqref{eqn:2d-phase}. The four phases---paramagnetic, toric code on edges, toric code on faces, and doubled toric code---have fixed points described by~\eqref{eqn:H-para}, \eqref{eqn:H-TCE}, \eqref{eqn:H-TCF}, and~\eqref{eqn:H-2TC}, respectively. The phase boundaries are vertical and horizontal lines because the edge qubits and face qubits are decoupled in the bulk.}
    \label{fig:2d-phase}
\end{figure}

Including more terms, as in 
\begin{align}
H = &- J_{Av} \sum_v A_v - J_{Ac} \sum_c A_c - J_{Zf} \sum_f Z_f \nonumber\\
& - J_{Xe} \sum_e X_e - J_{Be} \sum_e B_e - J_{Bf} \sum_f B_f, \label{eqn:2d-phase}
\end{align}
results in a larger phase diagram. Note that each term acts only on edges or only on faces, so that we can analyze each sector separately. The phases and phase transitions are shown in Fig.~\ref{fig:2d-phase}. The phase transitions happen at $J_{Bf} = J_{X_e}$ and $J_{Be} = J_{Zf}$. 

\begin{figure}
    \centering
    \includegraphics[width = \linewidth]{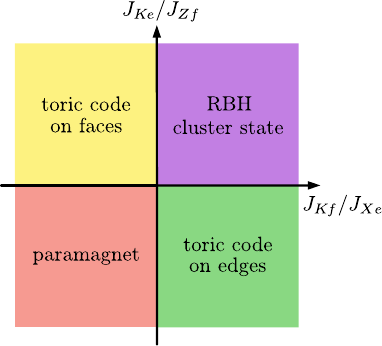}
    \caption{Phase diagram for the Hamiltonian in~\eqref{eqn:2d-phase-B}. The four phases---paramagnetic, toric code on edges, toric code on faces, and RBH cluster state---have fixed points described by~\eqref{eqn:H-para}, \eqref{eqn:H-TCE}, \eqref{eqn:H-TCF}, and~\eqref{eqn:H-RBH}, respectively. Analysis similar to Ref.~\cite{Muhlhauser2022} allows us to infer that the phase boundaries are vertical and horizontal lines. Compare to Fig.~6 of Ref.~\cite{LiPhaseDiagram}.}
    \label{fig:2d-phase-B}
\end{figure}

Another useful Hamiltonian is
\begin{align}
H = &- J_{Av} \sum_v A_v - J_{Ac} \sum_c A_c - J_{Zf} \sum_f Z_f \nonumber\\
& - J_{Xe} \sum_e X_e - J_{Ke} \sum_e K_e - J_{Kf} \sum_f K_f, \label{eqn:2d-phase-B}
\end{align}
which is the same as~\eqref{eqn:2d-phase} but with the $B$ terms replaced with $K$ terms. Using duality transformations like those in Ref.~\cite{Muhlhauser2022}, we can show that the phase transitions happen at $J_{Ke}=J_{Zf}$ and at $J_{Kf} = J_{Xe}$. The resulting phase diagram looks like Fig.~\ref{fig:2d-phase-B}. 
The full Hamiltonian $H_\ITC$ has a higher-dimensional phase diagram, with each of the five described phases appearing.

The analysis of the phase diagram suggests some possible connections to the KVC.
Reference~\cite{LiPhaseDiagram} shows that the KVC supports four phases: two 3d toric code phases and two trivial phases, where the trivial phases have 2d toric codes on opposite boundaries.
In the ITC, the paramagnetic and RBH phases are both trivial in the bulk, in the sense that neither is topologically ordered. However, they cannot be continuously connected without closing the gap or breaking the symmetry, meaning that $H_\RBH$ is in an SPT phase. Thus, we say that they are not the same phase but that both are trivial. Note that they have their 2d toric codes on opposite boundaries, as is true of the two trivial phase in Ref.~\cite{LiPhaseDiagram}. In this sense, $H_\Para$, $H_\RBH$, $H_\TCE$, and $H_\TCF$ are all found in the KVC. These phases arrange into a phase diagram (Fig.~6 of Ref.~\cite{LiPhaseDiagram}) that matches Fig.~\ref{fig:2d-phase-B} in the ITC.

However, the ITC also includes the doubled toric code phase ($H_\TTC$). Reference~\cite{LiPhaseDiagram} did not find such a phase in the KVC, but they did not exhaustively search the entire parameter space. If the KVC also contains a doubled toric code phase, then it would be possible that the KVC and ITC are related by some finite-depth unitary circuit, and that they are equivalent as subsystem codes.


\section{Remaining questions} \label{sec:questions}

In this paper we have introduced a single-shot quantum error-correcting code, the intertwined toric code. The ITC possesses three main advantages. First, the ITC is physically motivated  in that it descends directly from underlying toric codes and that the procedure that builds the ITC confines the problematic pointlike excitations of the toric codes. Second, the resulting code has geometrically simple logical operators and decoding procedures, both descending from the geometric simplicity of the toric codes. Third, the transparent nature of the check operators makes extracting the phase diagram for the ITC straightforward. 

This work uncovers a number of promising avenues for future research. 
In Sec.~\ref{sec:build} we constructed the ITC by adding single-site check operators to 3d toric codes, while in Sec.~\ref{sec:errors} we instead constructed it by adding single-site check operators to the RBH Hamiltonian. Both constructions suggest a generalization to arbitrary Walker-Wang models. The Walker-Wang construction~\cite{WalkerWang} takes a 2d anyon theory as input and outputs a 3d lattice model. Using two transparent bosons as the input theory gives two copies of the toric code~\cite{vonKeyserlingkSurfaceAnyons} while using the 2d toric code as input gives the RBH Hamiltonian~\cite{Roberts2020}. In both cases, the vertex terms of the Walker-Wang model are equivalent to the stabilizers in the ITC.
This suggests a procedure for turning a 2d anyon theory into a 3d subsystem code: Construct the Walker-Wang model corresponding to the anyon theory and then supplement with single-site check operators so that the vertex terms remain stabilizers but the face terms become check operators. Reference~\cite{BridgemanLifting} defines another procedure for constructing 3d subsystem codes from 2d anyon theories. How are these two procedures related? Could the subsystem Walker-Wang construction shed light on the differences between the GCC and the codes constructed in Ref.~\cite{BridgemanLifting}?

The RBH Hamiltonian also provides for fault-tolerant measurement-based quantum computation~\cite{RaussendorfFaultTolerant}, wherein a 2d toric-code degree of freedom is teleported across a 3d lattice system. Initially, the bulk of the system is in the ground state of the RBH Hamiltonian and the 2d toric code is on the left boundary. After measuring single-qubit operators in the bulk, the toric code ends up on the right boundary. In the language of the ITC, this process consists of using check operators to move from the paramagnetic phase to the RBH phase. In fact, single-shot error-correction procedures in the KVC correspond to scheduled phase transitions as well~\cite{LiPhaseDiagram}. What conclusions can be drawn from this connection between single-shot error correction and fault-tolerant measurement-based quantum computation?

One of the advantages of the KVC is that each check operator acts on at most three qubits. Is it possible to rewrite the ITC so that no check operator acts on more than three qubits? The presentation in Sec.~\ref{sec:build} uses smaller check operators. Is it possible to perform single-shot error correction using measurements of these check operators? Heuristically, it seems possible because a single measurement error introduces a geometrically close pair of violated stabilizers, rather than a single isolated violated stabilizer (as in the 2d toric code). However, this claim would have to be checked.

The boundaries of the ITC (and the GCC and the KVC) hold outsized importance. As originally introduced in Sec.~\ref{sub:bulk}, the ITC with no boundaries encodes no logical qubits. However, the analysis of Sec.~\ref{sec:errors} shows that we can reliably detect syndromes and correct errors even without a boundary. Taken together, these statements show that reliable error correction, which follows from confinement in the ITC, is insufficient for single-shot error correction of a logical qubit. What, then, is the physical interpretation of the boundary? Furthermore, the bare logical operators commute in the bulk and only anticommute on the boundary. In the phase diagrams of Sec.~\ref{sec:phase}, the boundaries ``host" the topological order when there is no topological order in the bulk (in the paramagnetic and RBH phases). The boundaries also ruin self-correction in the sense that models exist that are self-correcting in the presence of generic perturbations in the bulk but only with symmetry-allowed perturbations on the boundary~\cite{StahlSymmetry}. Clearly, the relationship between the bulk and the boundary contains interesting unanswered questions.

Lastly, all known single-shot stabilizer codes are self-correcting (but only exist in 4d or higher). None of the phases that we found in the ITC phase diagram are self-correcting. Can the simplified construction of the ITC offer any progress towards a new subsystem code that does include a self-correcting phase in 3d?

\section*{Acknowledgments}

I would like to thank Rahul Nandkishore for continued support and direction during this process. I also thank Yaodong Li and Marvin Qi for helpful discussion about the project, and Tyler Ellison, Aaron Friedman, Oliver Hart, Sam Roberts, David T. Stephen, and Matteo Wilczak for feedback on the manuscript. This work was supported by the U.S. Department of Energy, Office of Science, Basic Energy Sciences, under Award No. DE-SC0021346.

\appendix*
\renewcommand{\theequation}{A\arabic{equation}}
\section{More boundary conditions} \label{sec:appendix}

As promised, we here define the $e$-condensed and $m$-condensed boundary conditions so that the ITC can be defined with a single logical qubit on a cubic lattice that is itself a cube (with no periodic boundary conditions). 

We put the $e$-condensed boundary on the top and bottom of the cube. To do so, start with the previously-defined trivial boundary and remove qubits from the boundary edges. We are left with the same $B_e$ check operators but some $B_f$ check operators are truncated to 3-body operators, which we leave in the check group. Now there are no boundary $A_v$ stabilizers, but we keep the 5-body $A_c$ stabilizers. On this boundary, the bare logical membrane operator $\barelogical{Z}$ and the dressed logical stringlike operator $\barelogical{Z}'$ may both terminate,
\begin{align}
\inlinegraphic{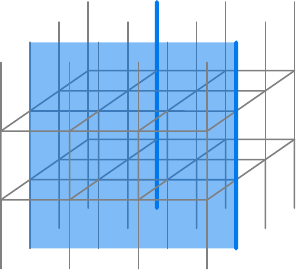},
\end{align}
but the $X$-type logical operators may not.

To construct the $m$-condensed boundary on the front and back of the cube, once again start with the trivial boundary. Now, add qubits to every boundary face. For every boundary edge we can now define a 3-body $B_e$ operator. We also define single-body $Z_f$ operators on these new face qubits. The resulting stabilizers are the 5-body $A_v$ operators we already had and new 6-body $A_c=\prod_{f\in \partial c} Z_f$ operators. The bare logical membrane operator $\barelogical{X}$ and the dressed logical stringlike operator $\barelogical{X}'$ may both end on this boundary,
\begin{align}
\inlinegraphic{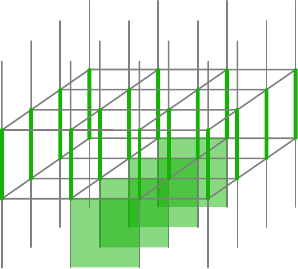},
\end{align}
but the $Z$-type logical operators may not.

These boundaries result in a cleaner code because they only encode a single logical qubit. There are $3L^3+2L^2-L$ edges and $3L^3-L^2$ faces for a total of $6L^3+L^2-L$ physical qubits. On these qubits we enforce $3L^3 - 3L$ single-site $X_e$ check operators, $2L^2 + 2L$ of the $K_e$ check operators (only counting those on the intertwined boundary), and $3L^3-L^2$ single-site $Z_f$ check operators. Between $3L^3 + L^2$ total $B_f$ check operators we have $L^3$ relations and therefore $2L^3+L^2$ independent $B_f$ operators. There are $3L^3-2L^2-L$ total $B_e$ check operators with $L^3-L^2-L+1$ relations between them, giving $2L^3-L^2-1$ independent $B_e$ operators. Along with $L^3+L^2-L-1$ of the $A_v$ stabilizers and $L^3$ of the $A_c$, we have 
\begin{align}
K = N - \frac{1}{2} \left( \log_2 |\G| + \log_2 |\S| \right) = 1
\end{align} 
logical qubit.

We should think of this as the minimal intertwined toric code because it only encodes a single logical qubit and has no need for periodic boundary conditions. We can go through the analysis of Sec.~\ref{sec:phase} to see what these boundaries become in the different phases. For example, the $e$-condensed boundary becomes the rough boundary conditions for the 2d toric code in the paramagnetic phase and for the 3d toric code in the TCE phase.

\bibliography{main}

\end{document}